\documentclass[onecollarge]{svjour2}
\usepackage{graphicx,amssymb,amsmath}
\journalname{Few-Body Systems}

\begin{document}

%\vspace*{-1.5cm}
%\begin{flushright}
%{SLAC-PUB-14212}\\
%\end{flushright}

\title{
% \begin{flushright}
% {\rm\small
% SLAC-PUB-14212}
% \end{flushright}
Hamiltonian Light-Front Field Theory: Recent Progress and Tantalizing Prospects}

\author{ J.~P.~Vary}
%\email{xbzhao, pmaris, jvary@iastate.edu}
\institute{J.~P.~Vary \at
Department of Physics and Astronomy, Iowa State University, Ames, Iowa 50011, USA, \email{jvary@iastate.edu}
}
%\author{ H.~Honkanen}
%\email{heli.m.honkanen@phys.jyu.fi}
%\institute{
%Department of Physics, University of Jyv\"askyl\"a, P.O. Box 35 (YFL), FI-40014 University of Jyv\"askyl\"a, Finland}
%\author{S.~J.~Brodsky}
%\email{sjbth@slac.stanford.edu}
%\institute{
%SLAC National Accelerator Laboratory, Stanford University, Menlo Park, California, USA }
\date{\today}
\maketitle 
\begin{abstract}
Fundamental theories, such as Quantum Electrodynamics (QED) and Quantum Chromodynamics (QCD) promise great predictive power addressing phenomena over vast scales from the microscopic to cosmic scales. However, new non-perturbative tools are required for physics to span from one scale to the next. I outline recent theoretical and computational progress to build these bridges and provide illustrative results
for Hamiltonian Light Front Field Theory. One key area is our development of basis function approaches that cast the theory as a Hamiltonian matrix problem while preserving a maximal set of symmetries. Regulating the theory with an external field that can be removed to obtain the continuum limit offers additional possibilities as seen in an application to the anomalous magnetic moment of the electron. Recent progress capitalizes on algorithm and computer developments for setting up and solving very large sparse matrix eigenvalue problems. Matrices with dimensions of 20 billion basis states are now solved on leadership-class computers for their low-lying eigenstates and eigenfunctions.
\end{abstract}

%%%%%%%%%%%%%%%%%%%%%%%%%%%%%%%%%%%%%%%%%%%%%%%%%%%%%%%%
\section{Introduction}
\label{intro}
%%%%%%%%%%%%%%%%%%%%%%%%%%%%%%%%%%%%%%%%%%%%%%%%%%%%%%%%
Quantum field theory in the non-perturbative domain presents opportunities 
and challenges to both particle physics and nuclear physics.  Increasingly, it
is also gaining attention in condensed matter physics.
Fundamental understanding of, among others, the phase structure of strongly interacting
systems, the spin structure of the proton, the neutron electromagnetic form factor, 
and the generalized parton distributions of the hadrons should emerge from results derived from a 
non-perturbative light-front Hamiltonian approach.
The light-front Hamiltonian, quantized within in a basis function approach as discussed here, 
offers a promising avenue that capitalizes on recent theoretical 
and computational achievements in quantum many-body theory.

For background, one notes that Hamiltonian light-front field theory 
in a discretized momentum basis \cite{Brodsky:1997de} and in transverse lattice
approaches \cite{JPV214,GP08} have shown significant promise.
I outline here a Hamiltonian basis function approach following Refs. \cite{Vary:2009gt,Honkanen:2010rc} 
that exploits recent advances in solving the non-relativistic strongly interacting nuclear
many-body problem \cite{NCSM,NCFC}.  There are many issues faced
in common that help one appreciate the need for sharing developments among subfields of physics 
such as how to: 
(1) define the Hamiltonian;
(2) regularize/renormalize for the available finite spaces while preserving all symmetries;
(3) solve for eigenvalues and eigenvectors;
(4) evaluate experimental observables; and,
(5) take the continuum limit.

I begin with a brief overview of recent progress in solving 
light nuclei with realistic nucleon-nucleon (NN) and three-nucleon (NNN)
interactions using {\it ab initio} no-core methods. 
I then outline a basis function approach suitable for light front gauge theories including the issues of renormalization/regularization. I present an outline of the approach to
cavity-mode QED, to QED in the absence of a cavity and sketch extensions to QCD. 
For a more detailed account of recent advances in applications to QED without a cavity, I refer to Ref. \cite{Zhao} in these proceedings.

%%%%%%%%%%%%%%%%%%%%%%%%%%%%%%%%%%%%%%%%%%%%%%%%%%%%%%%%
\section{No Core Shell Model (NCSM) and No Core Full Configuration (NCFC) methods}
\label{prog}
%%%%%%%%%%%%%%%%%%%%%%%%%%%%%%%%%%%%%%%%%%%%%%%%%%%%%%%%

To solve for the properties of self-bound strongly interacting systems, such as nuclei, with
realistic Hamiltonians, one faces immense theoretical and computational
challenges.  Recently, {\it ab initio} approaches have been developed that
preserve all the underlying symmetries and they 
converge to the exact result given sufficient computational effort.  The basis function approach 
\cite{NCSM,NCFC} is one of several methods shown to be successful.  
The primary advantages are its flexibility for choosing the Hamiltonian, the method
of renormalization/regularization and the basis space.  
These advantages support the adoption of the basis function approach 
in light-front quantum field theory.

Refs. \cite{NCSM,Hayes03,Shirokov07,MarisICCS10,chiral07,14C} and \cite{NCFC,Bogner07,Maris10} provide examples of the recent advances in the {\it ab initio} NCSM and NCFC, respectively.  The NCSM adopts 
a renormalization method that provides an effective interaction dependent on the chosen many-body basis space cutoff ($N_{\rm{max}}$ below). The NCFC either retains the un-renormalized interaction or adopts a basis-space independent renormalization so that the exact results are obtained either by using a sufficiently large basis space or by extrapolation to the infinite matrix limit.  Recent results for the NCSM employ realistic nucleon-nucleon (NN) and three-nucleon (NNN) interactions derived from chiral effective field theory to solve nuclei with Atomic numbers 10-13 \cite{chiral07}. For an overview of the NCSM including applications to reactions and to effective interactions with a core, see Ref.~\cite{NPN2011}. Recent results for the NCFC feature a realistic NN interaction that is sufficiently soft  that binding energies and spectra from a sequence of finite matrix solutions may be extrapolated to the infinite matrix limit \cite{Maris10}. Experimental binding energies, spectra, magnetic moments  and Gamow-Teller transition rates are well-reproduced in both the NCSM and NCFC approaches.  Convergence of long range operators such as electric quadrupole are more challenging.

It is important to note two recent analytical and technical advances. First, non-perturbative renormalization has been developed to accompany these basis-space methods and their success is impressive. Several schemes have emerged and current research focuses on understanding the scheme-dependence of convergence rates (different observables converge at different rates) \cite{Bogner07}. Second, large scale calculations are performed on leadership-class parallel computers to solve for the low-lying eigenstates and eigenvectors and to evaluate a suite of experimental observables.  Low-lying solutions for matrices of basis-space dimension about 20-billion on 215,000 cores with a 5-hour run is the current record.  However, one expects these limits to continue growing as the techniques are evolving rapidly \cite{MarisICCS10} and the computers are growing dramatically.  Matrices with dimensions in the several tens of billions will soon be solvable with strong interaction Hamiltonians.

In a NCSM or NCFC application, one adopts a 3D harmonic oscillator (HO) for all the particles in the nucleus (with HO energy $\hbar\Omega$), treats the neutrons and protons independently, and generates a many-fermion basis space that includes the lowest oscillator configurations as well as all those generated by allowing up to $N_{\rm{max}}$ oscillator quanta of excitations.  The single-particle states specify the orbital angular momentum projection and the basis is referred to as the $m$-scheme basis. For the NCSM one also selects a renormalization scheme linked to the basis truncation while in the NCFC the renormalization is either absent or of a type that retains the infinite matrix problem.  In the NCFC case \cite{NCFC}, one either proceeds to a sufficiently large basis that converged results are obtained (if that is computationally feasible) or extrapolates to the continuum limit.

%%%%%%%%%%%%%%%%%%%%%%%%%%%%%%%%%%%%%%%%%%%%%%%%%%%%%%%%
\section{Recent Results with the No Core Shell Model (NCSM) method}
\label{NCSM-results}
%%%%%%%%%%%%%%%%%%%%%%%%%%%%%%%%%%%%%%%%%%%%%%%%%%%%%%%%

The results of numerous {\it ab initio} NCSM applications not only show good convergence 
with regard to increasing size of the basis space but also have reproduced 
known properties of 0p-shell nuclei (nuclei up to $^{16}$O) as well as explain existing puzzles 
and make predictions of, as yet, unexplained nuclear phenomena. I cite a prominent example
to illustrate this point.

\begin{figure}[tb]
{\includegraphics[width=0.9\textwidth]{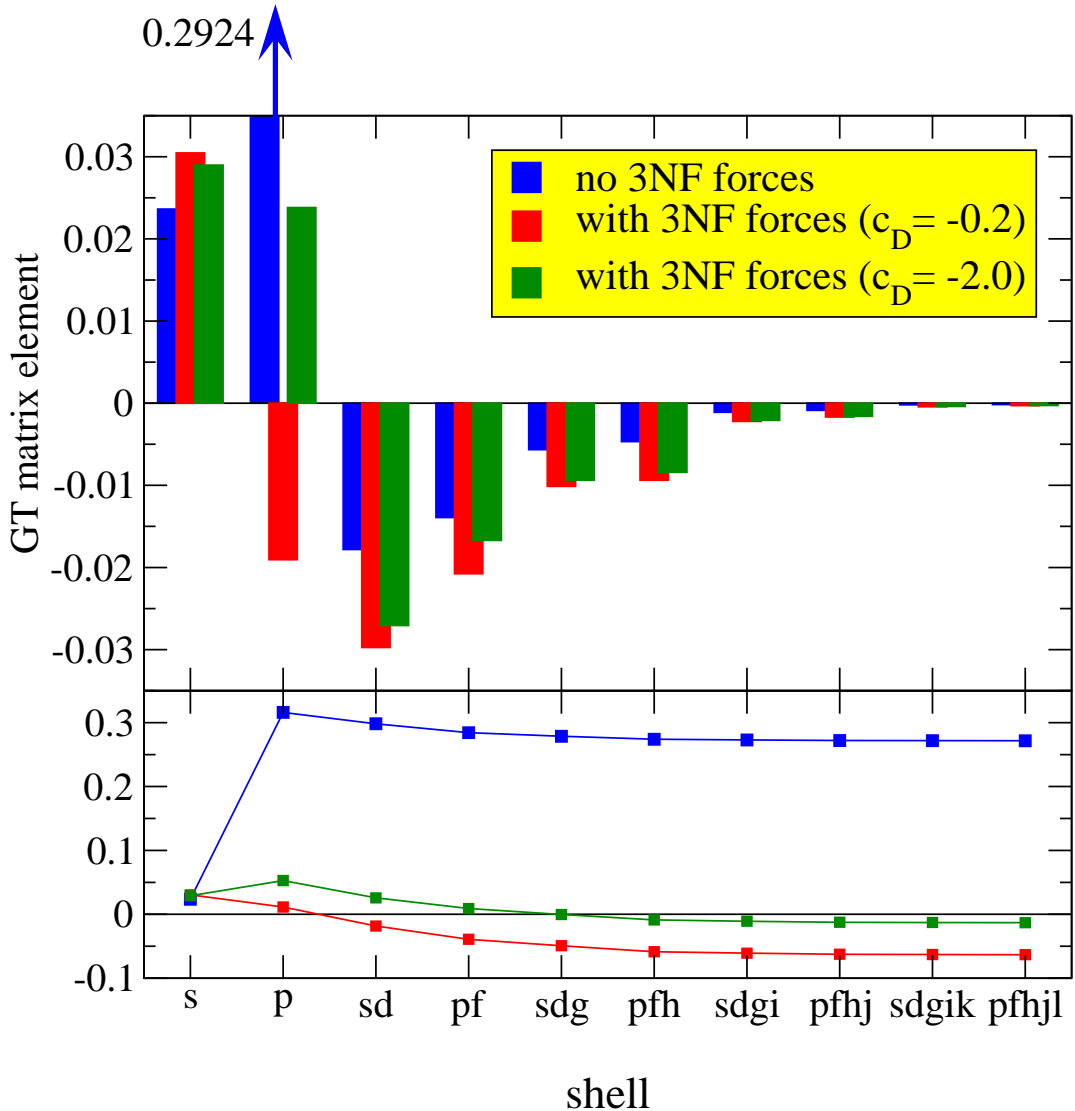}}
\caption{(Color online) Contributions to the $^{14}$C beta decay matrix element as a function of the 3D harmonic oscillator shell in the basis space when the nuclear structure is described by the $\chi$EFT interaction (adopted from Ref. \cite{14C}). Top panel displays the contributions with (two right bars, the red and green, of each triplet) and without (leftmost bar, the blue bar, of each triplet) the NNN force at $N_{\rm max} = 8$. Contributions are summed within each shell to yield a total for that shell. The bottom panel displays the running sum of the GT contributions over the shells with the same color coding scheme. Two reasonable choices for coupling constants (red and green components of the histogram and lines) in the NNN-interaction lead to similar strong suppression of the GT matrix element. Note, in particular, the 
order-of-magnitude suppression of the $0p$-shell contributions arising from the NNN force.}
\label{C14MGT}
\end{figure}

A recent calculation has evaluated the Gamow-Teller (GT) matrix element for the beta decay 
of $^{14}$C, including the effect of chiral NNN forces~\cite{14C}. These investigations show that the very long lifetime for $^{14}$C arises from a cancellation between 0p-shell NN-and NNN-interaction contributions to the GT matrix element, as shown in Figure \ref{C14MGT}. The net result is a GT matrix element close to zero (final point of the green curve in the lower half of Fig. \ref{C14MGT}) which is far more consistent with the 5730 year halflife of 
$^{14}$C. The same calculations show that including the NNN-interactions also bring the binding energies of $^{14}$C and $^{14}$N into closer agreement with experiment. These 
A=14 beta decay results were obtained in the largest basis space achieved to
date with NNN interactions, $N_{\rm{max}}=8$, or approximately one billion m-scheme configurations.

Other noteworthy results include recent calculations for $^{12}$C
explaining the measured $^{12}$C B(M1) transition from the g.s. to the $(1^+,1)$ state at 15.11 MeV 
and showing more than a factor of 2 enhancement arising from the NNN interaction.  
Neutrino elastic and inelastic cross sections on $^{12}$C were shown to be similarly 
sensitive to the NNN interaction and their contributions significantly improve 
agreement with experiment~\cite{Hayes03}.  Working in collaboration 
with experimentalists, we uncovered a puzzle in the GT-excited 
state strengths in A=14 nuclei~\cite{Negret}.  Its resolution may lie in the 
role of intruder-state admixtures, but this will require further work.

%%%%%%%%%%%%%%%%%%%%%%%%%%%%%%%%%%%%%%%%%%%%%%%%%%%%%%%%
\section{Light-front Hamiltonian field theory}
\label{LF}
%%%%%%%%%%%%%%%%%%%%%%%%%%%%%%%%%%%%%%%%%%%%%%%%%%%%%%%%

It has long been known that light-front Hamiltonian quantum field theory has similarities with non-relativistic quantum many-body theory and this has prompted applications with established non-relativistic many-body methods (see Ref. \cite{Brodsky:1997de} for a review).  These applications include theories in 1+1, 2+1 and 3+1 dimensions.  Several of my efforts in 1+1 dimensions, in collaboration with others, have focused on developing an understanding of how one detects and characterizes transitional phenomena in the Hamiltonian approach.  To this end, I list the following developments:

\begin{enumerate}
\item{identification and characterization of the quantum kink solutions in the broken symmetry phase of  two dimensional $\phi^4$ including the extraction of the vacuum energy and kink mass that compare well with classical and semi - classical results  \cite{Chakrabarti:2003ha};}
\item{detailed investigation of the strong coupling region of the topological sector of the two-dimensional $\phi^4$ theory demonstrating that low-lying states with periodic boundary conditions above the transition coupling are dominantly kink-antikink coherent states \cite{Chakrabarti:2003tc};}
\item{ switching to anti-periodic boundary conditions in the strong coupling region of the topological sector of the two-dimensional $\phi^4$ theory and demonstrating that low-lying states above the critical coupling are dominantly kink-antikink-kink states as well as presenting evidence for the onset of kink condensation\cite{Chakrabarti:2005zy}.}
\end{enumerate}

More recently, full-fledged applications to gauge theories in 3+1 dimensions have appeared and there are additional talks at this conference showing initial results for QED (e.g. by Hiller and by Chabysheva) plus roadmaps for addressing QCD (e.g. talk by Glazek).  A brief summary of some of the major developments in 3+1 dimensional Hamiltonian light front field theory includes the solutions of:

\begin{enumerate}
\item{light-front QED wave equations for the electron plus electron-photon system~\cite{Hiller,Chabysheva:2009vm,Chabysheva:2009ez}}
\item{simplified gauge theories with a transverse lattice~\cite{JPV214,GP08,Dalley:2002nj}}
\item{Hamiltonian QED for the electron plus electron-photon system in a trap with a basis 
function approach~\cite{Vary:2009gt,Honkanen:2010rc} and without a trap~\cite{Zhao,Zhao:2011}.}
\end{enumerate}

These successes open pathways for ambitious research programs to evaluate non-perturbative amplitudes and to address the multitude of experimental phenomena that are conveniently evaluated in a light-front quantized approach.  As one important example, consider the deeply virtual Compton scattering (DVCS) process which provides the opportunity to study the 3-dimensional coordinate space structure of the hadrons.  Recent efforts with model 3+1 dimensional light-front amplitudes \cite{Brodsky:2006in,Brodsky:2006ku} have shown that the Fourier spectra of DVCS should reveal telltale diffractive patterns indicating detailed properties of the coordinate space structure. As another major example consider the non-perturbative regime of QED that future experiments with ultra-strong pulsed lasers will explore, for example, looking for non-perturbative lepton pair production~\cite{Adare:2009qk,Ruf:2009zz,Dumlu:2010ua,Ilderton:2010wr} (see talk by Ilderton).  Another application resides with the strong time-dependent QED fields generated in relativistic heavy-ion collisions~\cite{Tuchin:2010gx}.

%%%%%%%%%%%%%%%%%%%%%%%%%%%%%%
\section{Basis Light-Front Quantization (BLFQ) approach}
\label{BLFQ}
%%%%%%%%%%%%%%%%%%%%%%%%%%%%%%

Together with co-authors, I have introduced the ``Basis Light Front Quantized'' (BLFQ) approach 
\cite{Vary:2009gt} which adopts a light-front basis space consisting of the 2D harmonic oscillator (HO)  for the transverse modes (radial coordinate $\rho$ and polar angle $\phi$) and a discretized momentum space basis for the longitudinal modes with either periodic or anti-periodic boundary conditions.  The 2D oscillator states are chosen to retain rotational symmetry about $x^-$ direction and they are characterized by their principal quantum number $n$ and orbital quantum number $m$.  Adoption of this basis is also consistent with recent developments in AdS/QCD correspondence with QCD \cite{deTeramond:2008ht,Brodsky}. Note, however, that the choice of basis functions is arbitrary except for the standard conditions of orthonormality and completeness.

In our initial applications, we focus on QED and consider a system including only $|e\rangle$ and $|e\gamma\rangle$ sectors in a transverse scalar harmonic trap  \cite{Honkanen:2010rc} and, more recently, in the absence of the external trap \cite{Zhao,Zhao:2011}.  Both of these setups, once the Fock space is extended, will be useful for addressing a range of strong field QED problems such as electron-positron pair production in relativistic heavy-ion collisions and with ultra-strong pulsed lasers planned for the future.  We adopt the sector dependent non-perturbative renormalization scheme \cite{Karmanov:2008br}.

The chosen basis allows the imposition of symmetry constraints that reduce the Hamiltonian matrix dimension considerably for a desired precision.  For example, we impose the constraint of a fixed total magnetic projection ($J_z$) and fixed total longitudinal momentum in dimensionless units ($K$) consistent with longitudinal boost invariance.  We also impose a cutoff in the Fock space basis controlling the number of fermion and boson degrees of freedom and we impose a limit on the maximum total 2D oscillator quanta 
($N_{\rm{max}}$) in the basis.  We then investigate how the non-perturbative results depend on the cutoffs and seek to obtain results in the continuum limit where the cutoffs are removed.

The total light-front Hamiltonian is $H=H_0+V$ ($KH$ gives the invariant mass-squared) where the unperturbed Hamiltonian $H_0$ for this system  is defined by the sum of the occupied modes with the scale set by the combined constant $\Lambda^2 = 2M_0\Omega$ with $\Omega$ representing the HO frequency of the trap:

\begin{eqnarray}
&&  H_0 = 2M_0 P^-_c 
= \frac{\Lambda^2}{K}\sum_i{\frac{2n_i+|m_i| +1 +
{{\bar m_i}^2}/\Lambda^2
}{x_i}},
\label{Hamiltonian}
\end{eqnarray}
where ${\bar m_i}$ is the mass of the parton $i$ and $x_i$ is its light-front momentum fraction. 
We keep the photon mass set to zero and the electron
mass $\bar m_e$ is set at the physical mass 0.511 MeV in our 
non-renormalized calculations. We also set  $M_0=\bar m_e$.  More recently, we remove the transverse trap and simply replace $H_0$ with the pure kinetic energy operator in the transverse center of momentum frame \cite{Zhao,Zhao:2011}. 

The interaction vertices defining $V$ in Eq. \ref{Hamiltonian} are taken directly from light-front quantized QED in the light-front gauge and are used to generate the interaction matrix elements in the basis.  Considerable analytical and numerical efforts are required to achieve an efficient and accurate evaluation scheme for these matrix elements.

In the most recent application to QED, we still retain the truncated basis including only $|e\rangle$ and $|e\gamma\rangle$ sectors as in Ref.~\cite{Honkanen:2010rc}. However, we introduce major extensions and improvements. For a more complete description, I refer to the paper by Zhao at this meeting \cite{Zhao} and to a separate paper~\cite{Zhao:2011}. Here, I simply list a few of the key extensions and improvements.\\

1. In order to expand the range of applications, we extend the application of BLFQ to a free space system by omitting the external transverse trap. \\

2. In order to improve computational efficiency and numerical precision, we replace numerical integrations previously used in Ref.~\cite{Honkanen:2010rc} to evaluate matrix elements of QED interaction vertices with newly-developed analytic methods.\\

3. To achieve improved convergence, we allow the HO basis length scale to be fixed separately in each Fock sector which allows a more efficient treatment of the transverse center-of-momentum degree of freedom.\\

4. We correct the evaluation of the anomalous magnetic moment $a_e$ and a factor appearing in the vertex matrix elements.  These corrections go in opposing directions for the previously evaluated $a_e$ in an
external trap~\cite{Honkanen:2010rc} and updated results will be provided in a separate paper~\cite{Zhao:2011}.\\

%%%%%%%%%%%%%%%%%%%%%%%%%%%%%%%%%%%%%%%%%%%%%%%%%%%%%%%%
\section{Results for Electron Anomalous Magnetic Moment $a_e$}
\label{result}
%%%%%%%%%%%%%%%%%%%%%%%%%%%%%%%%%%%%%%%%%%%%%%%%%%%%%%%%
With the extensions and improvements summarized in Sec.~\ref{BLFQ}, we evaluate and diagonalize the light-front QED Hamiltonian in $|e\rangle$ and $|e\gamma\rangle$ sectors and evaluate $a_e$ from the resulting light-front amplitude for the lowest mass eigenstate. In this work we reduce the QED coupling constant $\alpha$ by a factor of $10^4$ in order to reduce higher order effects and facilitate comparison with $a_e$ from perturbation theory~\cite{Schwinger:1948iu}. In addition, we omit the instantaneous electron exchange vertex for the same reason.

We define our basis space with total longitudinal momentum $K$=80 which we found adequate for the present application but will be extended in the future.  In fact, the results presented in Ref.~\cite{Zhao} already extend the basis to $K$=160.  Furthermore, we use 2D HO single-particle states with frequencies $\omega$ ranging from 0.01MeV to 1.4MeV.  These $\omega$'s bracket the electron mass $m_e$=0.511 MeV, the only scale-setting parameter in the QED Hamiltonian. At each $\omega$ we calculate $a_e$ with $N_{\rm{max}}$ in the range of 10 to 118 to map out its convergence behavior with increasing $N_{\rm{max}}$. Larger $N_{\rm{max}}$ translates to a larger basis with higher effective ultraviolet cutoff and lower effective infrared cutoff in the transverse plane. We expect that, with increasing $N_{\rm{max}}$, the results more closely approximate the Schwinger result. The rate of  convergence may be different for different $\omega$'s, depending on $a_e$'s sensitivity to the effective cutoffs of the basis space. Our results agree with this expectation and approach the Schwinger result uniformly as $N_{\rm{max}}$ increases with increments of 4.

\begin{figure}[!t]
\centering
\includegraphics[width=0.9\textwidth]{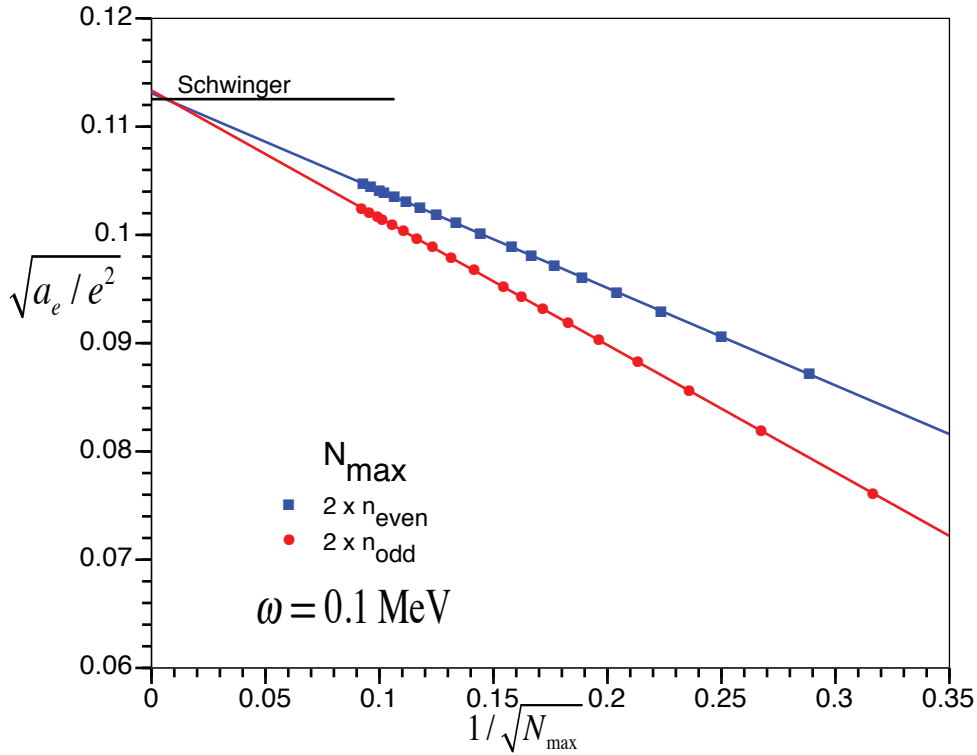}
\caption{(Color online) Anomalous magnetic moment of the electron calculated in BLFQ compared to the Schwinger result~\cite{Schwinger:1948iu}. The vertical axis is the square root of anomalous magnetic moment normalized to electron charge, $e$, so the Schwinger value is $\sqrt{\frac{1}{8\pi^2}}=0.11254$. The horizontal axis is the square root of the reciprocal of $N_{\rm{max}}$. Symbols are for the BLFQ results. Squares: even $N_{\rm{max}}$/2; circles: odd $N_{\rm{max}}$/2. The lines are linear extrapolations of BLFQ results based on all the points shown which span $N_{\rm{max}}=10-118$}
%at or above $N_{\rm{max}}=64$.}
\label{anoma}
\end{figure}

In Fig.~\ref{anoma} I present the results evaluated with $\omega$=0.1MeV.  For comparison, see the results in Ref.~\cite{Zhao} at $\omega$=0.02MeV and $\omega$=0.5MeV.
For each $\omega$ the results exhibit a simple pattern with increasing $N_{\rm{max}}$: the results with even $N_{\rm{max}}/2$ are systematically larger than those with odd $N_{\rm{max}}/2$ so that the former and the latter separate into two individual groups. Within each group the results define a trend which is understandable by analysis of the perturbative calculation in light-front QED~\cite{Zhao,Zhao:2011,Brodsky}.  Other features of these results are similarly understandable~\cite{Zhao,Zhao:2011}.

The data points in Fig.~\ref{anoma} appear to define straight lines as a function of $1/\sqrt{N_{\rm{max}}}$ as can be seen by the linear fits to all the points shown (solid lines). We can therefore easily extrapolate to the limit of no basis truncation  ($N_{\rm{max}}\to\infty$) where we expect to recover the Schwinger result. Indeed as seen in Fig.~\ref{anoma} the lines converge close to the Schwinger result in this limit. Their intercepts at 1/$N_{\rm{max}}$=0 are: 0.1131(1.0\%) and 0.1133(1.4\%) for even $N_{\rm{max}}/2$ and odd $N_{\rm{max}}/2$, respectively. The percentages in the parenthesis are their corresponding relative deviation from the Schwinger result, $\frac{a_e}{e^2}=\frac{\alpha}{2\pi e^2}$=$\frac{1}{8\pi^2}\approx$0.012665.  

What is not so apparent from a visual inspection of Fig.~\ref{anoma} is the fact that the extrapolated values comes closer to the Schwinger result if one limits the linear fit to results for only the larger values of $N_{\rm{max}}$.  For example, if the linear fit is performed for $N_{\rm{max}}\ge64$ the extrapolated values improve to
0.1129(0.7\%) and 0.1130(0.9\%) for even $N_{\rm{max}}/2$ and odd $N_{\rm{max}}/2$, respectively.  Continuing this avenue of investigation, if the linear fit is performed only for results with $N_{\rm{max}}\ge100$ the extrapolated values improve to 0.1128(0.4\%) and 0.1129(0.6\%) for even $N_{\rm{max}}/2$ and odd $N_{\rm{max}}/2$, respectively.  This is an encouraging sign of expected systematic improvement with increasing 
$N_{\rm{max}}$.

What is also important to note is that these results are systematically improvable.  We will extend the calculations to larger $K$ and $N_{\rm{max}}$ values to further improve accuracy and reduce extrapolation uncertainties.  That is we will evaluate additional results in regions where they are expected to scale more accurately as a function of $\sqrt{\frac{1}{N_{\rm{max}}}}$. In order to compare with the perturbative result for $a_e$ with the  rescaling as shown in Fig.~\ref{anoma} (i.e. to achieve results for $\frac{a_e}{e^2}$) it is also advantageous to further decrease the fine structure constant below $10^{-4}\alpha$, the value for the results presented here. 
  
%%%%%%%%%%%%%%%%%%%%%%%%%%%%%%%%%%%%%%%%%%%%%%%%%%%%%%%%
\section{Conclusions and Outlook}
\label{end}
%%%%%%%%%%%%%%%%%%%%%%%%%%%%%%%%%%%%%%%%%%%%%%%%%%%%%%%%

The recent history of light-front Hamiltonian field theory features many advances that pave the way for non-perturbative solutions of gauge theories.   The goal is to evaluate the light-front amplitudes for strongly interacting composite systems and predict experimental observables.  High precision tests of the Standard Model may be envisioned as well as applications to theories beyond the Standard Model.

We can extend the BLFQ approach to QCD by implementing the SU(3) color degree of freedom for each parton - 3 colors for each fermion and 8 for each boson.   We have investigated two methods for implementing the global color singlet constraint.   In the first case, we follow Ref. \cite{Lloyd} by constraining all color components to have zero color projection and adding a Lagrange multiplier term to the Hamiltonian to select global color singlet eigenstates.   In the second case, we restrict the basis space to global color singlets \cite{Vary:2009gt,JunLi}.  The second method produces a factor of 30-40 lower many-parton basis space dimension at the cost of increased computation time for matrix elements. Either implementation provides an exact treatment of the global color symmetry constraint but the use of the second method provides overall more efficient use of computational resources. Nevertheless, the computational requirements of this approach are substantial, and we foresee extensive use of leadership-class computers to obtain practical results.

I would like close by mentioning that we are extending the QED application in several directions. One specific goal is to include the capability to treat strong time-dependent laser pulses to address non-perturbative QED processes~\cite{Ilderton:2010wr} such as those presented at this meeting (see talk by Ilderton).  In addition, we are launching an initial effort to evaluate the properties of charmonium in a BLFQ treatment of QCD. 

\vspace{5 mm}

The author thanks all his collaborators on the cited publications as well as K.~Tuchin, J.~Hiller, S.~Chabysheva, V.~Karmanov, A.~Ilderton, Y. Li and P. Wiecki for fruitful discussions.  
Computational resources were provided by the National Energy Research Scientific Computing Center (NERSC), which is supported by the Office of Science of the U.S. Department of Energy under Contract No. DE-AC02-05CH11231.
This work was supported in part by US DOE Grants DE-FG02-87ER40371 and  DE-FC02-09ER41582 (UNEDF SciDAC Collaboration).
This work was also supported in part by US NSF grant 0904782.

%\vspace{-.3cm}

\end{document}